\documentclass[final,12pt]{elsarticle}
\usepackage{graphicx}
\usepackage{amssymb}
\journal{Electoral Studies}

\begin{document}
\begin{frontmatter}



\title{On the reliability of voting processes: the Mexican case}
\author[UAM-A]{G. B\'aez}
\author[UAM-Acb]{H.~Hern\'andez-Salda\~na}
\author[ICF-UNAM]{R.A.~M\'endez-S\'anchez}

\address[UAM-A]{Lab. de Sistemas Din\'amicos, Dpto. de Ciencias B\'asicas,
Universidad Aut\'onoma Metropolitana-Azcapotzalco,
Av. San Pablo 180, 02200, M\'{e}xico D.F., Mexico.}
\address[UAM-Acb]{Dpto. de Ciencias B\'asicas,
Universidad Aut\'onoma Metropolitana at Azcapotzalco,
Av. San Pablo 180, 02200, M\'{e}xico D.F., Mexico.}
\address[ICF-UNAM]{Instituto de Ciencias F\'{\i}sicas, Universidad Nacional Aut\'onoma de M\'exico,
A.P 48-3, 62210, Cuernavaca, Mor., M\'exico.}

\begin{abstract}
Analysis of vote distributions using current tools from statistical physics is of increasing interest.
While data considered for physics studies are subject to a careful understanding of 
error sources, such analysis are almost absent in studies of voting process.
As a way to test  reliability in electoral processes we choose to investigate 
the July 2006 presidential election in Mexico.
We use records which appeared in the Programa de Resultados Preliminares, PREP, the 
program which offers electoral results as soon as they are captured.
In order to contrast the results we used the congressional election data-set and 
the July 2000 presidential record. 
We demonstrate the existence of correlations in the percentage of votes
and, we offer evidence of the strong influence of the PRI in the curious
results in the presidential case.  
Distributions of errors in the data-sets are  analyzed for all the 
elections and no large deviations were found. That is, even when
the sum of errors is around 45\% 
in all cases,
their global behavior is similar. This result 
gives support to the thesis of no large fraud for the July 2006 presidential 
election. 
Distribution of votes in all cases is obtained for all 
the parties. Parties and candidates with few votes, annulled votes 
and non-registered candidates follow a power law distribution, 
and the corporate party follows a daisy model distribution. 
Parties and candidates with many votes show a mixed 
behavior.

\end{abstract}

\begin{keyword}
vote distribution \sep election  \sep opinion dynamics 
\sep error analysis \sep election forensics
\PACS 87.23.Ge \sep 89.75.-k

\end{keyword}

\end{frontmatter}

%   IMPORTANT: DO NOT ERASE
%   KEYS
% P1 PAN
% P2 PRI  Alianza por M\'exico PRI + PVEM
% P3 PRD  Coalici\'on por el bien de todos PRD+ PT + P Convergencia 
% P4 ROBERTO CAMPA, NUEVA ALIANZA
% P5 PATRICIA MERCADO, ALTERNATIVA

% For the 2000 process
% Alianza por el Cambio, FOX, PAN and PVEM
% PRI
% Alianza por Mexico, C. Cardenas, PRD, PT, Convergencia, PAS, Partido de la Sociedad Nacionalista. 
% PCD
% PARM
% DS
% Non registered
% Annuled 

\section{Introduction}

Elections are an important matter for humanity. Analysis of how we vote 
is an important subject in political, social and economical sciences. In recent year, physicists
and mathematicians became interested in looking if some of the dynamics and universal behaviors 
found in statistical mechanics,  complex systems and other fields can be used to 
understand the voting process and the way we form our opinion. Such an area has its own name: sociophysics.
The book of Ph. Ball~\cite{Ball} offers a clear and readable introduction to the physicists' and mathematicians' 
incursion in social and economical subjects, meanwhile the report ~\cite{CastellanoReport} is a much more 
technical approach.

%The study of how we vote and several aspects of its dynamics
%has been of interest in the last years for the physicist community~\cite{Ball,Borghesi,Castellano,CostaFilho1999}, increased by the 
%large amount of data currently available on the Internet. 

In this context, several theoretical models have been successfully applied in order to explain 
some characteristics on how we take decisions and, in particular, how we vote. 
However,  studies of actual electoral data with  the current tools in physics are 
scarce~\cite{Borghesi, Borghesi2010, Borghesi2012a,Borghesi2012b,CostaFilho2006}
\cite{Castellano2007,HernandezSaldanaE1,Herrmann}.
The current work is part of an effort to fulfill such an absence and contribute to 
characterize the statistical properties of actual databases.
Certainly, all these approaches are compl\-imen\-tary to the vast literature
and studies done in political, social, economical and anthropological sciences.

Here, we are interested in a particular kind of electoral forensics 
related with the reliability of vote processes. 
The faithfulness of the data  is an important issue,
since the record of social processes is subject to all kinds of 
uncontrollable factors. Furthermore, while data considered for 
analysis in physics  are subject to a careful understanding of 
the sources of error, such analysis are almost absent in 
studies of voting process even though election forensics is 
a current subject in political sciences.

%Such studies are complementary to the large amount of social, political and anthropological 
%literature and mainly has  been focused in modelling the opinion formation. 

%tools of statistical mechanics, complex systems and
%other fields of physics and mathematics have been used to
%understand the voting process as part of the so called sociophysics~\cite{Borghesi,Castellano}.

%In an increasing number of cases, 
%the advent of new technologies is changing the 
%way in which elections are realized, not only the way that  
%information  appears in the mass media, but also the 
%emergence of real-time data. Public access to partial results 
%at the same time that votes are  counted could be the rule in the
%future. In many processes the results are clear; 
%however, in closed elections, some counterintuitive results appear.
%Understanding this behavior is important, since our experience with 
%{\em uncorrelated} situations give us an insight that could be 
%wrong when we deal with {\em correlated} variables.

In this work, we present an approach to two problems: we analyze  the real time results 
obtained during the federal election in Mexico in the year of 2006, and 
we offer an analysis of the distribution of errors obtainable 
from this public database. Both problems contribute to clarify 
the dynamics in the Mexican electoral processes, where a misunderstanding 
of the data give place to suspicious of the existence of a large fraud (Mega-fraud). 
We present statistical arguments that limit such a suspicious, 
gives a better perspective of the peculiarities of the Mexican
processes where the nonexistence of widespread fraud does not necessarily mean absence
of irregularities. We offer, as well, evidence that usual statistical 
assumptions  are not necessarily fulfilled by the electoral processes.
The reported forensics
in the peer reviewed literature does not consider the
tests analyzed here. On the other hand, we do not consider the 
pre-election events and how they influenced the final result.

We focus the present study on data obtained from the ``Program of Preliminary Electoral Results'', or PREP, after 
its acronym in Spanish~\cite{PREP}, which is the  system implemented by the electoral authorities
(Instituto Federal Electoral, IFE) in order to present the first results as the information arrives to the 
headquarters.
In Mexico, the election is performed using electoral cabins (polling stations) that 
admit, by construction, around $750$ voters and that are, approximately, uniformly distributed
over the population with the right to emit a vote.
The PREP works with 
certificates stamped on the packets of ballots (named electoral packets). 
In those certificates the citizens who staff polling places write, by hand,
the number of votes received for each party, total number of those votes, 
etc., at the end of the electoral day. Later, the authorities of each 
electoral cabin deliver the electoral packets and their certificates to the capture centers.
The time of arrival is captured as well as the results stamped on the electoral packets.
The records are sent to Mexico City headquarters and
published on the electoral authorities' website~\cite{IFEweb}. The 
final results are recorded in a public file~\cite{PREP}. The analysis 
presented here is based on such a file. 
During the next days IFE admits and discusses objections and appellations. The final results
are published in the final count named  ``Conteo Distrital'' or District Count. 
Disputes about electoral practices are solved by the autonomous Federal Electoral Tribunal. 
Analysis on statistical aspects of distrital count results are in progress and some have been published 
\cite{HernandezSaldanaE1,HernandezSaldanaE2,HernandezSaldanaE3}.

Federal elections in Mexico are performed each six years on the first Sunday of July,
including presidential change and renewal of both chambers. For details on how the Mexican electoral  
systems works see the note of Klesner~\cite{Klesner}. An important result is  the 
conformation of the chambers, that depends in a complicated way on the 
alliances and the relative strength of each political current, see Crespo for an 
explanation~\cite{Crespo11}.

The political parties who participated in the July 2006 election are grouped in two alliances and several parties.
The first party is {\it Partido Acci\'on Nacional} (PAN) postulating as candidate to Felipe Calder\'on Hinojosa (FCH); 
the first alliance  corresponds to {\it Alianza por M\'exico} composed of the {\it Partido Revolucionario Institucional} (PRI)
and  the {\it Partido Verde Ecologista de M\'exico} (PVEM), they postulated Roberto Madrazo Pintado (RMP);
the second alliance was {\it Alianza por el bien de todos} composed of the
following political parties: {\it Partido de la Revoluci\'on Democr\'atica} (PRD),
{\it Partido del Trabajo} (PT) and {\it Convergencia}, with candidate  
Andr\'es Manuel L\'opez Obrador (AMLO);
the last parties  are {\it Partido Nueva Alianza} with Roberto Campa Cifri\'an (RCC) 
and,  {\it Partido Alternativa} to Patricia Mercado Castro (PMC). The certificates report votes for 
non registered candidates and annulled votes as well.  
For the alliances we shall use the main party to identify them, that is
we shall call PRI instead of {\it Alianza por M\'exico} and PRD instead of
{\it Alianza por el bien de todos}.

The rest of the  paper is organized as follows: In the next section, we 
clarify our lines of argument. We establish the working hypothesis and
the alternative one. In section 3, we analyze correlations in the real-time
behavior of PREP data. For section \ref{Section:ConservationLaw}, we perform an analysis on 
the six independent self-consistent error distributions that can 
be build up with the PREP data. As bonus we describe the vote distribution 
for all the cases in section \ref{Section:Distributions}. Conclusions appear in section 6 and some additional information in an appendix.   
 
%show the time behavior of the record
%that PREP yields and we analyze the reliability of the system. Using
%some simple conservation laws (self-consistency), in 
%Section~\ref{Section:ConservationLaw}, we make the error analysis of
%data which appear in the PREP file.
%In Section~\ref{Section:Distributions} we show and analyze the 
%distributions of votes for  deputies, senators 
%and president. A brief conclusion about the error of the election 
%using the results of the PREP follows.

\section{The argumentative line}

The Mexican presidential election in 2006 was controversial and suspicions 
of a large (Mega) fraud has been present since election day. This 
was at the origin of our interest in analyzing electoral 
data. However, our lack of expertise in electoral matters shall rule  the 
statistical approach we use. Even when we have experience analyzing
statistical properties of complex physical systems, we do not have  a clear
idea of how a Mega-fraud in elections must looks like. Under this premise, 
a serious working hypothesis is to depart from the idea that no fraud 
exists and to perform as many test as we are able, in order to find a qualitative
difference in the results of all the cases we shall present. 

As pointed out by Good and Harding \cite{Good2004}, it is a source of error
to ``[use] the same set of data to formulate hypothesis and then to 
test those hypothesis''. Since the data set under suspicion is the presidential
election of 2006, we shall contrast all our test with the chambers election of 
the same year. We complement our results with the analysis of presidential election 
in Mexico during 2000.

Hence, our working hypothesis is established as: \\ 
\\
$H_1:$ {\sl Data from PREP for the Mexican elections of 2006, presidential and both chambers, 
present the same qualitative statistical behavior.} 
\\

The alternative hypothesis is: \\ 
\\
$H_0:$ {\sl Data from PREP for the Mexican elections of 2006, presidential and both chambers,
present a qualitative different statistical behavior.}
\\

This means, that manipulation in a large scale is present in one of the cases. The meaning
of ``qualitative difference'' will be clear in the section of error distributions.
Notice that these hypothesis do not depend on what we expect 
or not, according to our prejudices. For instance, for physicists it is 
natural to expect error distributions with a Gaussian or a Lorentzian shape. 
But, in the present case, we are not dealing with the same objects (particles)
or interactions. So, deviations from the canonical, uncorrelated,  behavior 
do not to be assumed as a signature of fraud when appear in different processes. 

The philosophy is that before we look what is out of place, we look, first, what is 
in place. Hence, our study is of a phenomenological nature.

\section{Time behavior of PREP}

Around $20:00$ hrs on July 2nd, 2006, the {\it Program of 
Preliminary Electoral Results} (PREP) of IFE~\cite{PREP} 
began to publish on its website the vote percentages for the presidential election of that day. 
The update was every $5$ minutes and the news services reproduced it. 
A very close election, 
according to opinion polls, was expected between the two main candidates 
of PAN and PRD. (Recall we shall use the names of the main political parties in a coalition).
At the beginning, the tendency showed a decrease in the number of votes for the PAN's candidate
and an increase for the candidate of PRD. 
At first sight, a crossing seemed to be imminent between the percentage of votes obtained by the two candidates. 
For this reason, several scientists as well as laymen started the capture of
real-time data using different methods, some captured by hand, 
others captured automatically with programs like {\em perl} 
\cite{Mochan,Baqueiro}.

\begin{figure}[t!]
\includegraphics[width=\columnwidth]{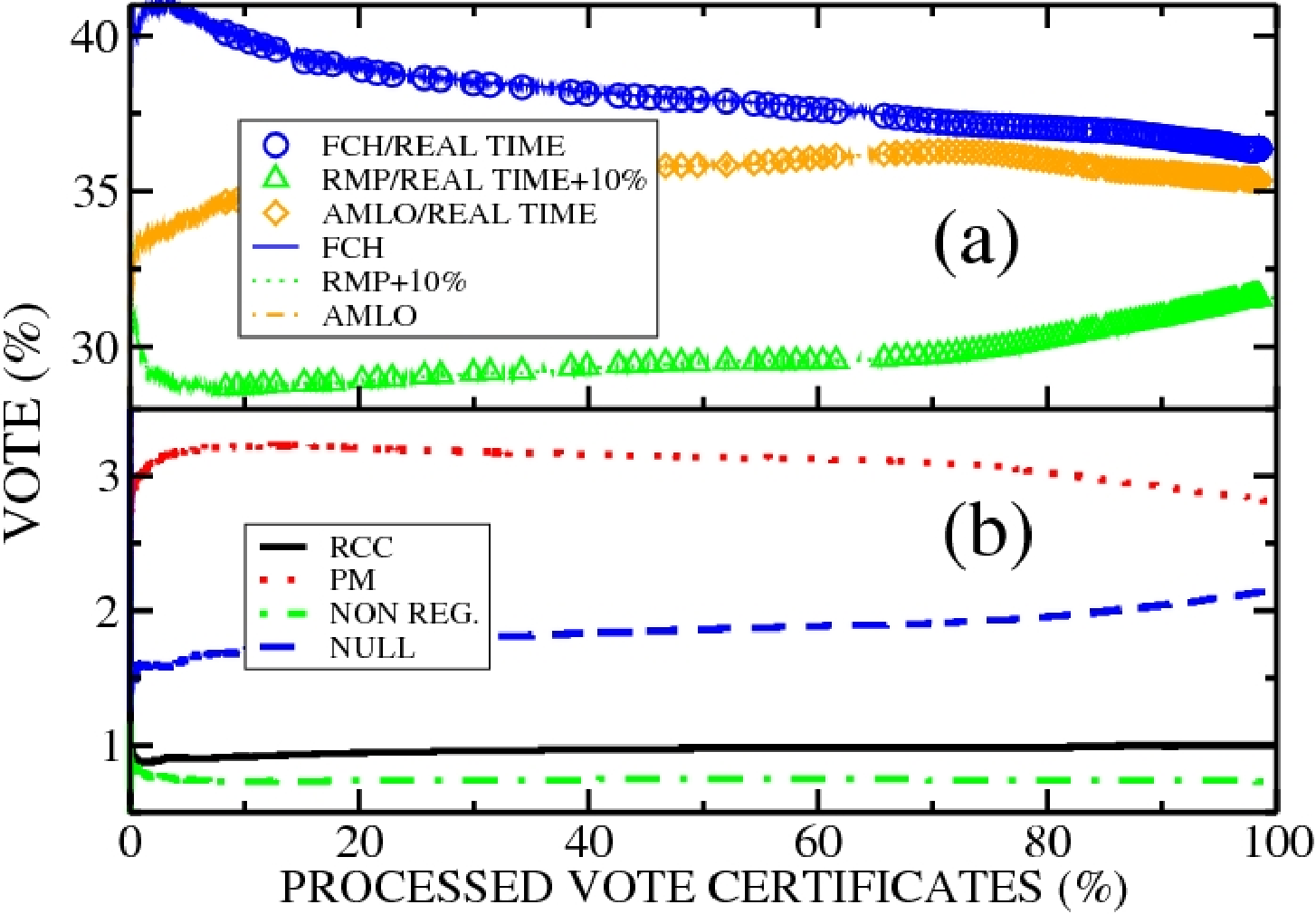}
\caption{(Color on-line) Real-time data given by the Program of Preliminary Electoral 
Results (PREP) for (a) the three main presidential candidates and (b) the other two candidates, the votes for non registered candidates and the annulled votes (NULL). 
The plots in (a) show the percentage of votes 
obtained per candidate against the percentage of processed vote certificates. 
The blue circles correspond to Felipe Calder\'on Hinojosa (FCH), candidate of PAN, 
the green triangles to Roberto Madrazo Pintado (RMP) for PRI, and the orange diamonds to Andr\'es Manuel L\'opez Obrador for the PRD party.
The data were taken 
from the Internet \cite{Mochan,Baqueiro} and correspond to the real time data published by 
IFE's web page. We added 10\% of votes to RPM in order to show all the 
results in a smaller window. Notice the change in the slope of this party after the 
$70 \%$ of processed certificates. We show the complete data from the PREP file\cite{PREP} as well, they are 
indicated by 
the corresponding lines.  From the same data set we plot, in (b), the percentages obtained for Roberto Campa Cifri\'an (RCC) for PANAL, Patricia Mercado for Partido Alternativa, we included the vote for non registered  candidates (NON REG.) and annulled votes (NULL). Notice that the change in the percentages, for these cases, is small.} 
\label{Fig:RealTime}
\end{figure}

In Fig.~\ref{Fig:RealTime}(a) we show the plot of the real-time data 
given by the PREP for the three main candidates captured in election day~\cite{Baqueiro}
as well as the data obtained from the PREP file (in lines at the figure). This plot shows 
roughly two tendencies for the percentage obtained by the candidate of PRD. 
The increasing tendency changed to a decreasing one around 
3:00 AM (around $ 70\%$ of the processed vote certificates in Fig.~\ref{Fig:RealTime}) of July 3rd.
As can be seen in the same figure, the PRI candidate increased its 
vote percentage changing the tendencies of the candidates of PAN and PRD, respectively.

%\newpage

%\newpage 

As a result, no crossing between the vote percentage of the two main 
candidates was found with the PREP in real time. In order to test the 
suspicious that real time data published in the web page of IFE is not coincident
with the reported data in the PREP file, we plotted both of them in 
Figure~\ref{Fig:RealTime}(a), and they agree. 

A common belief is to consider that  the behavior of a graph like 
Figure~\ref{Fig:RealTime} reaches its final value in a direct way, 
and a change like the presented here looks suspicious. However,
one must be careful, since we cannot consider {\it a priori uncorrelated} behavior. 
The graph shows {\it vote percentage} and,
evidently, this fact introduces correlations between the variables as we shall explain. 
For the sake of clarity we drop the multiplication by $100$.
The percentage of votes for party 1, for instance, at certain time $t$ or percentage of
processed vote certificates is calculated by 
\begin{equation}
\% V_1 (t) = \frac{V_1(t) }{\sum_i^N V_i(t)}, %  V_1+V_2+\cdots + V_N},
\end{equation}
and for party 2,
\begin{equation}
\% V_2(t) = \frac{V_2(t)}{\sum_i^N V_i(t)}. %  V_1+V_2+\cdots + V_N},
\end{equation}
Here $V_i(t)$ corresponds to the number of votes received by party $i$ at time $t$.
Eliminating the common expression $\sum_i^N V_i(t)$ between both equations we have
\begin{equation}
\% V_1 (t) = \frac{V_1(t)}{V_2(t)} \times \% V_2(t).
\end{equation}
This correlates the percentage $\% V_1 (t)$ with the precentage $\% V_2 (t)$, 
not mattering that 
the variables $V_1(t)$ and $V_2(t)$ are random or not. The same is true for the 
rest of the percentages and for all times. 
Additionally, the percentage of all the parties must sum $100\%$ as well, hence, it is natural 
that a ``mirror'' effect appears in normalized quantities. In  
Figure~\ref{Fig:RealTime}(b) we show the percentage of vote for the rest of the 
parties, notice that the values do not present significant changes, hence, the 
changes are due to the {\em three} main parties.

Moreover, the 
temporal data are not taken from a {\em uniform} sample and 
so we do not expect a fast convergence to the final results. 
Assuming a clean election, the first data that arrived to the 
capture centers were from sites with better transport networks and/or a better
vote counting performance. In Ref.~\cite{Pliego}, correlation 
with an official marginalization index~\cite{marginalizationindex} is presented and confirms this assertion: 
Reports from polling stations inside regions with a low marginalization index arrived
early to capture centers. 
In an earlier version of this paper\footnote{The first version of this paper was made public the 
13/Sept/2006, at the arxiv site \texttt{http://arxiv.org/abs/phys/0609114}. This site offers preprints in a 
free way and it is maintained by the University of Cornell.} 
we presented examples of how the 
sampling of polling stations  rules the shape of the percentage vote graph~\cite{HernandezSaldanaE0}. 
Evidently, the final values are the same.

The question that remains is: Does it show some peculiar behavior the {\em number} of votes?
A way to answer it is to calculate the correlation matrix, and particularly the auto-correlation.
In Table~\ref{tabla1} we report both, the correlations for percentage of votes in the 
lower triangular matrix and the correlations for the number votes in the upper triangular matrix.  
Recall that correlation matrices are symmetric. In the diagonal we report the result of auto-correlations
for percentage of votes, the corresponding auto-correlation in number of vote are equal to one in all the 
cases.

For the lower triangular matrix, the correlation in percentage  of votes shows 
the obvious relation between them, the vast majority is far from zero and near one or 
minus one. 
An exception in the case of the presidential  candidate
of PANAL, RCC,  whose auto-correlation is $0.08$(i.e. it is not auto-correlated!). Other anomalies appear with this party and
are reported below. 
Note that not all the auto-correlations in Table~\ref{tabla1} are unity, 
since the percentage and its momenta are calculated prior to calculating each correlation and a 
large variability in the PREP 
records exists. These statistics are reported in section \ref{Section:ConservationLaw}. 

The correlation matrix of percentage of votes for deputies and senators election are  presented in the 
appendix in Tables \ref{tabla3} and \ref{tabla4}. The results are similar to those presented in \ref{tabla1}: percentage
of votes are correlated (or anti-correlated).  An interesting point is that PANAL's percentage of vote auto-correlation is $0.71$ and $0.62$ for deputies and senators, far from the $0.08$ for presidential election.

For the number of votes, as expected, the story is different. The vast majority of correlations reported are close to zero and  all the auto-correlations of votes are one. Even the correlation of PANAL 
with other parties is around zero. The values calculated for the chambers are all consistent with this result: 
There is no a qualitative difference in the presidential and chambers elections.

A test on the statistical significance of the correlation coefficients 
can be done using a re-sampling method~\cite{Good2005}. However we found much more instructive 
to show the behavior of the  errors in the database. Which is the matter of the next section.

\begin{table}

\begin{center}
\begin{tabular}{|l||r|r|r|r|r|r|r|}
\hline
      &   PAN    & PRI    & PRD     &  PANAL & Alter & N.R.  & N.V.  \\ \hline \hline
PAN   &   1.00   &  -0.05 & -0.23   & 0.07  & 0.35    & 0.08  &  -0.06 \\ %\hline
PRI   &  -0.77   &    1.00 &  -0.15   & 0.10  & -0.17 &  0.07    & 0.18  \\ %\hline
PRD   & -0.74    &  0.15   &    1.02  & -0.02 & 0.34  &  0.01    &  0.00 \\ %\hline
PANAL & -0.26    &  0.17   &   0.22  &    0.08  &  0.12   & 0.07  & 0.08   \\ %\hline
Alter &  0.70    & -0.96   &   -0.07 & -0.16    &   1.00  &   0.12   & -0.06  \\ %\hline
N.R. & 0.11   &   0.06  & -0.25 &     -0.03  & -0.04   & 0.82   & 0.03  \\ %\hline
N.V. &  -0.95 &    0.81  &     0.62 &  0.25  &   -0.78 & -0.19  & 1.00 \\ \hline

\end{tabular}
\protect\caption{Correlation matrix of {\it vote percentage}, lower triangular matrix, 
and {\it number of votes}, upper triangular, for the presidential election in 2006. 
Diagonal corresponds to auto-correlation of vote percentage. Notice the anomalous 
value of the auto-correlation in the PANAL case. N.R. stands for non registered candidates and N.V. for
annulled votes.\label{tabla1}}
\end{center}
%\label{tabla1}
\end{table}

%\begin{table}
%\begin{tabular}{||c||r|r|r|r|r|r|r|}
%\hline \hline
% & P1 & P2 & P3 & P4 & P5 & I.C. & An.  \\
%\hline
%\hline

%P1 &   1.0000  &           &            &           &          &        &            \\
%P2 & -0.7734   &    1.0000 &            &           &          &        &             \\
%P3 &-0.7461   &   0.1461  &     1.0208 &           &          &        &             \\
%P4 &-0.2572   &    0.1708 &     0.2205 &   0.0786 &          &        &             \\ 
%P5 & 0.6982   &  -0.9594  &   -0.0754  &  -0.1618  &   1.0035 &        &             \\
%I.C. & 0.0462   &   0.0429  & -0.1234    & -0.0092   & -0.0034  & 0.7969 &             \\ 
%An. & -0.9492   &   0.8000  &    0.6294  &    0.2476 &  -0.7781 & -0.1455&        1.0011  \\  
%
% CORRELATION OF VOTES
%1.0000    &            &            &            &         &        &                  \\
% -0.0528  &  1.0000    &            &            &         &        &                  \\    
% -0.2343  &    -0.1516 &     1.0000 &            &         &        &                  \\
%  0.0744 &  0.0954  & -0.0160   & 1.0000     &         &        &                   \\ 
%  0.3480  &    -0.1734 &    0.3357  &   0.1213   &  1.0000 &        &                 \\
%  0.0817  & 0.6517    & 0.0142    & 0.0662    & 0.1217  &  1.0000&                  \\
% -0.0610  &  0.1789    &  -0.0032   &  0.07825   & -0.0626 &  0.0298&    1.0000    \\
% \hline
%\end{tabular}
%\caption{Correlation matrix of {/taki
%it vote percentage}. I.C. stands for independent candidates and An. for 
%annulled votes.}
%\label{tabla1}
%\end{table}

\begin{figure}[t!]
\includegraphics[width=0.9\columnwidth,angle=0]{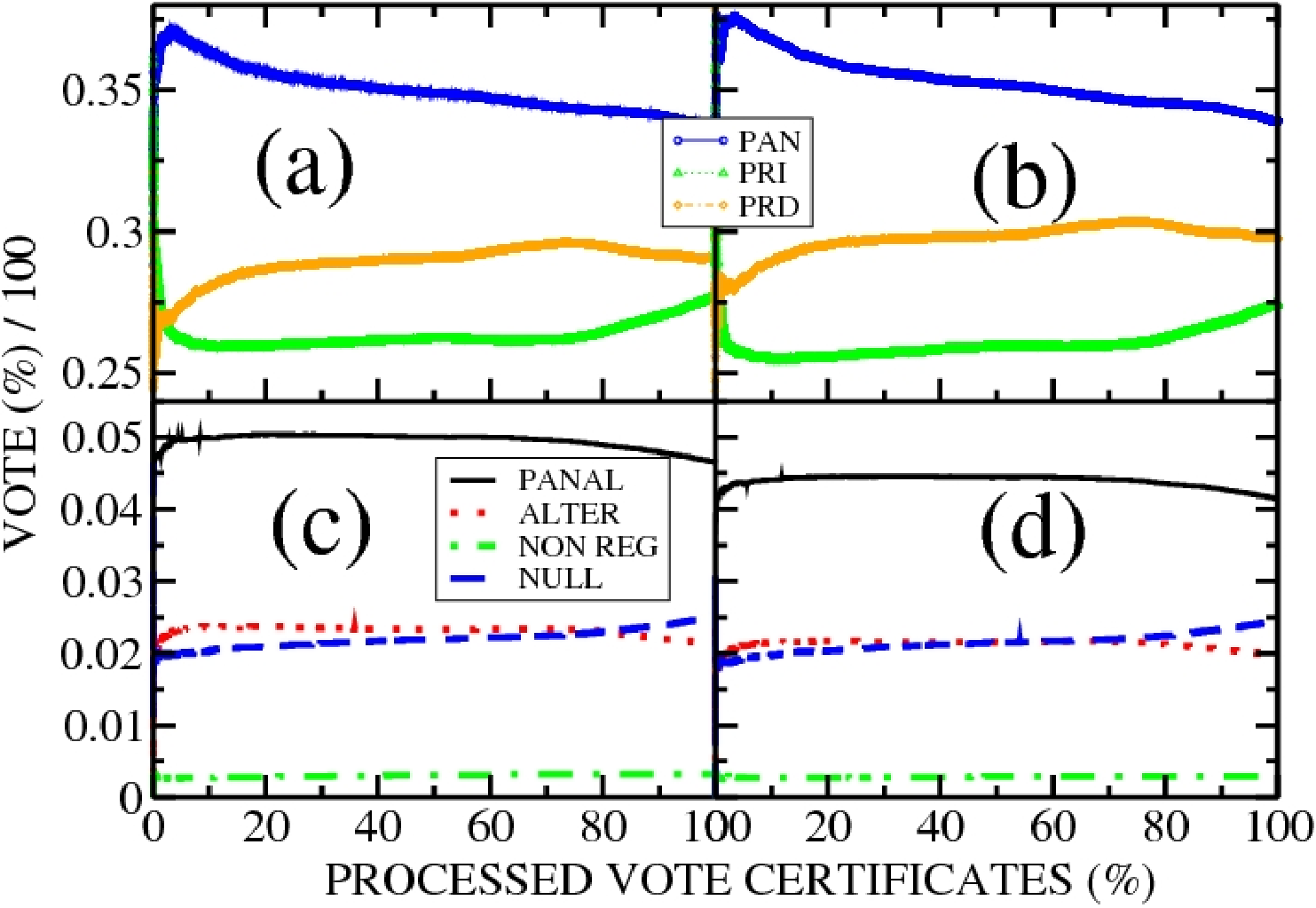}
\caption{(Color on-line) 
Same as the previous figure but for election of both chambers. Here the data were obtained from  PREP file and 
we do not have the results published in real time in IFE's web page.
In (a) percentage of votes for the three main parties/coalitions for the deputies election and in (b) for the 
high chamber. Except for the scales the behavior is similar to the  presidential case. A shift (not shown) 
of the percentage of votes for PAN and PRD for both chambers 
present the same non-crossing effect as in previous figure. 
Notice, as well, that PRI have a revival around the $70\%$ of processeed vote certificates in both cases.
In (c) and (d) we plot the results for the rest of the  parties, non registered and annulled votes for 
deputies and senators, respectively.
Notice that the global behavior in (a) and (b) is similar to figure~\ref{Fig:RealTime}(a). 
That is: vote for the chambers follows the presidential behavior.  
}
\label{Fig:Chambers2006}
\end{figure}

Returning to the percentage of votes behavior, we analyze the deputies and 
senators behavior. In Figure ~\ref{Fig:Chambers2006} we report the equivalent plots of Fig.~\ref{Fig:RealTime} for the Chambers. In Fig.~\ref{Fig:Chambers2006}(a) and(c) for deputies meanwhile in (b) and (d) for Senators. The behavior reported for 
the presidential data is similar in the current case, except that the apparent crossover is not present since the percentages of vote are different. For the 2000 election the behavior is similar and the results for the three main candidates 
is presented in Fig. ~\ref{Fig:PREP2000} at the appendix. It is interesting that a shift (not shown) on the data reproduce the
global behavior in the 2006 case, i.e. the growth of the PRI percentage of vote around $70\%$ of vote certificates and the falling of 
percentage of votes for PRD and PAN before that. 

\section{Conservation laws in elections}\label{Section:ConservationLaw}

\begin{figure}[t!]
\includegraphics[width=1.0\columnwidth]{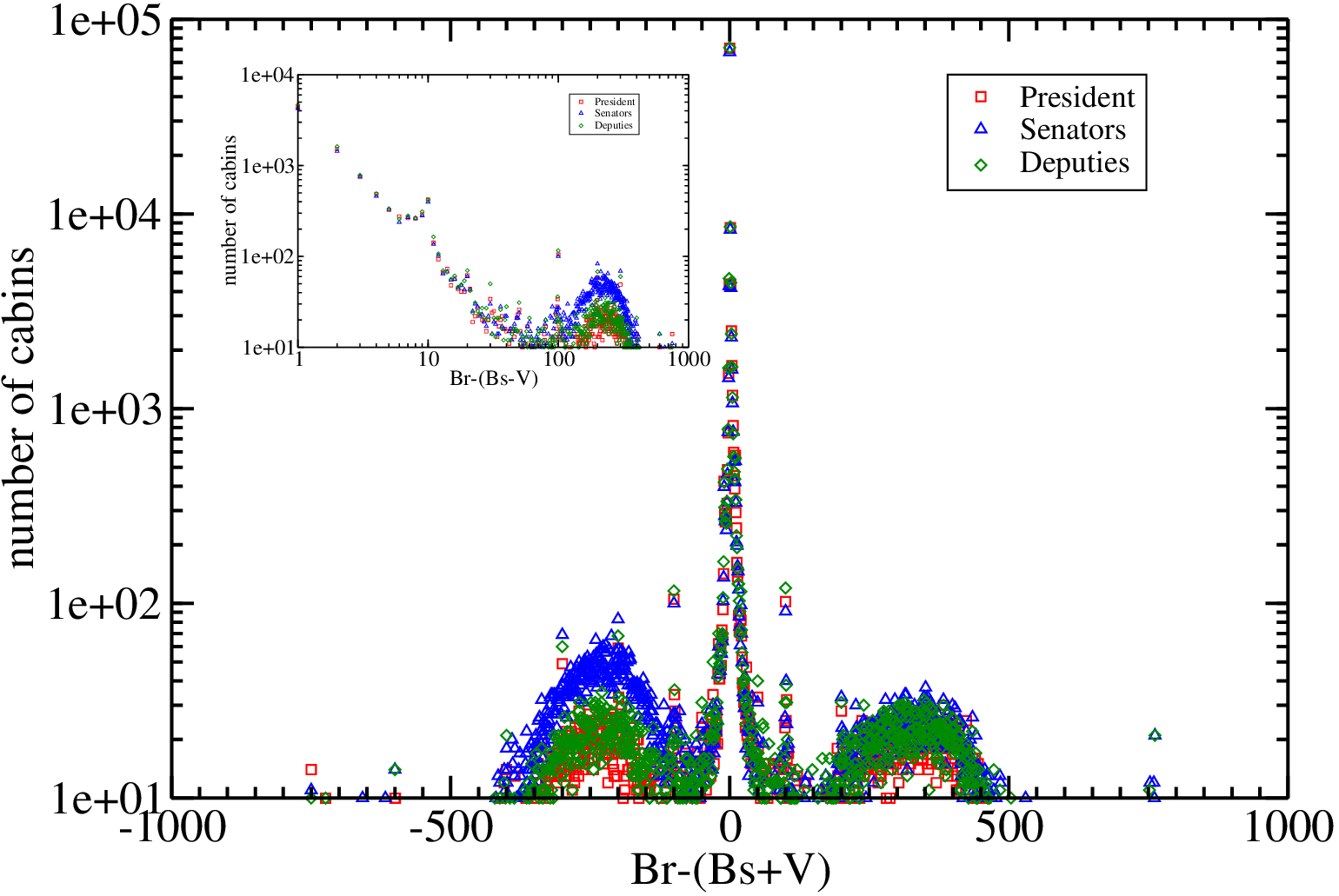}
\caption{(Color on-line) Error distributions, in logarithmic scale, of the difference between the total ballots 
received Br and the sum of the remaining ballots Bs and the total 
number of voters V per cabin, E$_1 = $ Br-(Bs+V), for all the cases in July 2006 election. 
Negative values in the horizontal axis 
mean more votes than ballots, i.e. there were more voters than documents which certified this fact, 
and positive values mean appearance of ballots,
but in both cases there is no conservation of the total number of 
received ballots Br per cabin. The inset shows the left branch of the 
distribution in log-log scale. This shows a power law decay. Note 
the several sharp peaks in the distribution along the horizontal axis 
between the values $10$ to $100$. 
We assume that these peaks are related with typos, but further analysis is required.
We note that the distribution for senators 
show a slightly different behavior in its left branch, compared with the corresponding  
distributions for president and deputies.
The distributions are not normalized.}
\label{Fig:LostBallots}
\end{figure}

Another way to test the reliability of the electoral processes is 
to check the self-consistency of the data-base.
In the cabin certificates several data are recorded and several 
of them must agree. For instance the total number of ballots received the
election day must be equal to the number of the remaining ballots plus the used 
ones. In an ideal election such kind of quantities must give zero for all the
cabin records, however we are dealing with an actual process performed 
by people and some errors could occur, intentional or not. In many processes Gaussian or lorentzian
distributions of error appear but in the current case we have no a priori 
knowledge about the distribution or if the whole amount of votes are enough to 
reach the limit distribution, if it exists. So, as explained in the argumentative 
line section, we define and explore the errors that could give us 
self-consistency in the database and contrast all of them.

The PREP database is composed from the data stamped in the electoral package by 
the citizens sorted by IFE to conduct the election at the polling station, 
several of them are easily subject to human error or to intentional alteration and, its distributions  must 
be object of study. The data are: Total number of 
received ballots at the beginning of the electoral process (Br), number of 
remaining (not used) ballots (Bs), number of voters (V), number of 
deposited ballots per cabin (Bd) and the number of votes received for each
party/candidate (V$_i$, $i=$ PAN, PRI, PRD, PANAL, ALTENATIVA, 
non-registered candidates and annulled votes). 
The conservation laws that we checked, for 
self-consistency, are six and they are summarized in 
table ~\ref{tabla2}.

\begin{table}

\begin{center}
\begin{tabular}{|l|l|c|}
\hline

E$_1$& B. received $-$ (B. not used $+$ Number of voters)& Br - (Bs + V) \\ 

E$_2$& B. received $-$ (B. not used $+$ B. deposited)& Br - (Bs + Bd) \\ 
E$_3$& B. received $-$ (B. not used $+$ Votes for each party)& Br - (Bs+$\sum_i$ V$_i$) \\ 
E$_4$& Number of voters $-$  B. deposited
& V - Bd  \\ 
E$_5$& Number of voters $-$ Votes for each party& V - $\sum_i$ V$_i$ \\ 
E$_6$& B. deposited $-$ Votes for each party & Bd - $\sum_i$ V$_i$ \\ 
\hline

\end{tabular}
\protect\caption{\label{tabla2}Table of errors considered for a self consistency test of the 
PREP database. The quantities considered are
those that were calculated and registered independently by the electoral
authorities (citizens) when
the votes were counted. In this sense, they are non-trivial. We use B. as a
shorthand for Ballots. The variable $i$ stands for the number of votes obtained
for each party/candidate.
}
\end{center}
\end{table}

The error distribution is build up by calculating the error E$_k$  
on each cabin and, then, make the histogram of the values obtained, i.e., how many cabins have
values of E$_k$ equal to 0,1,2,$\cdots$. Certainly the values can be negatives as well.  
In the current work we do not normalize the distributions.

\begin{figure}[t!]
\includegraphics[width=1.0\columnwidth]{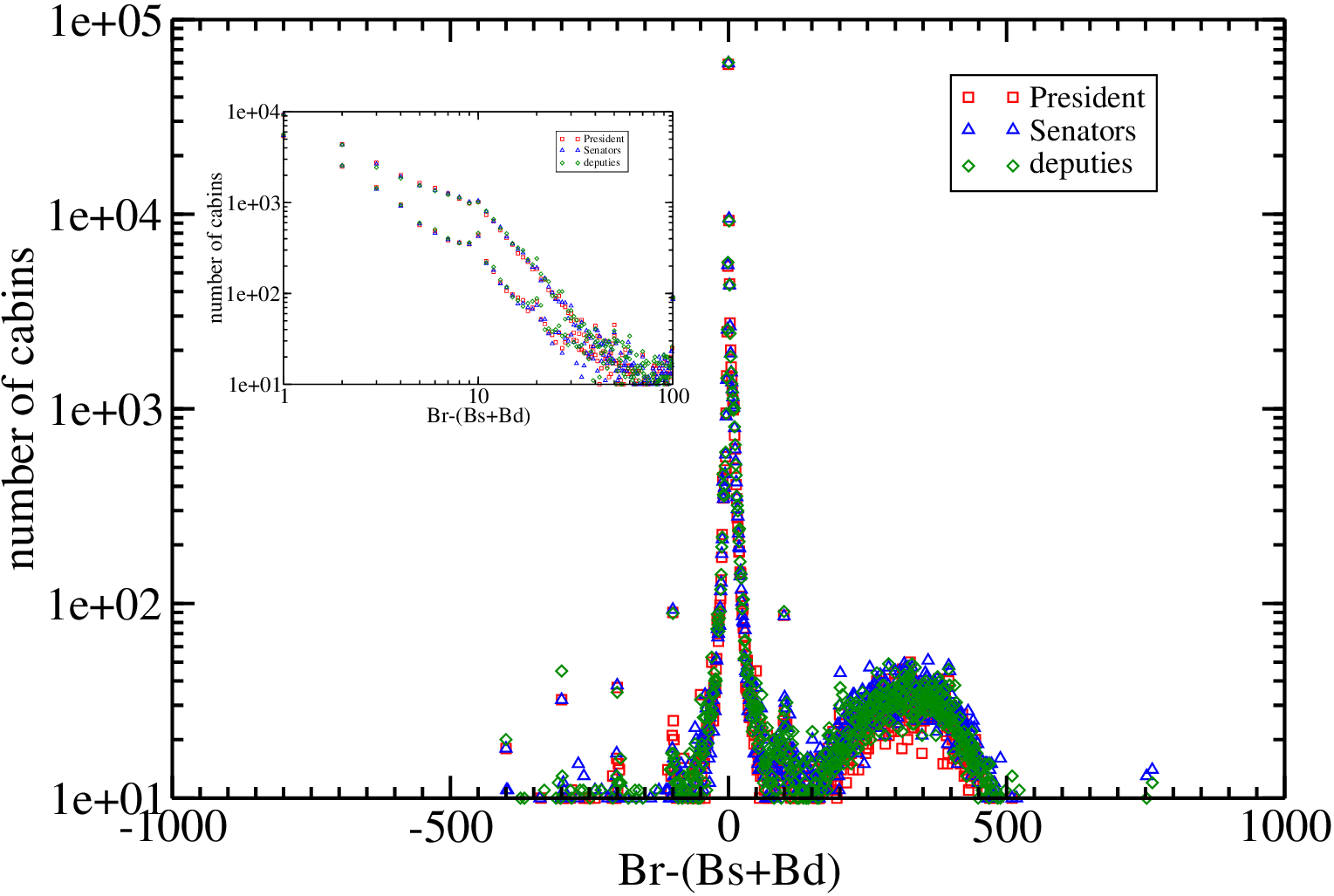}
\caption{(Color on-line) Error distributions, in logarithmic scale, of the difference between the total 
ballots received Br minus the sum of the ballots remained Bs plus the 
ballots deposited in urns Bd by cabin, i.e., E$_2 =$ Br-(Bs+Bd). 
Notice that the asymmetry between the both branches for the three 
distributions is similar, but it is different from that in previous figure.
Again, the left branch of this probability distributions shows the sharp peaks we
associated to {\it capture mistakes}. The inset shows both branches of the distributions in log-log scale.} 
\label{Fig:LostDeposited}
\end{figure}

\begin{figure}[t!]
\includegraphics[width=1.0\columnwidth]{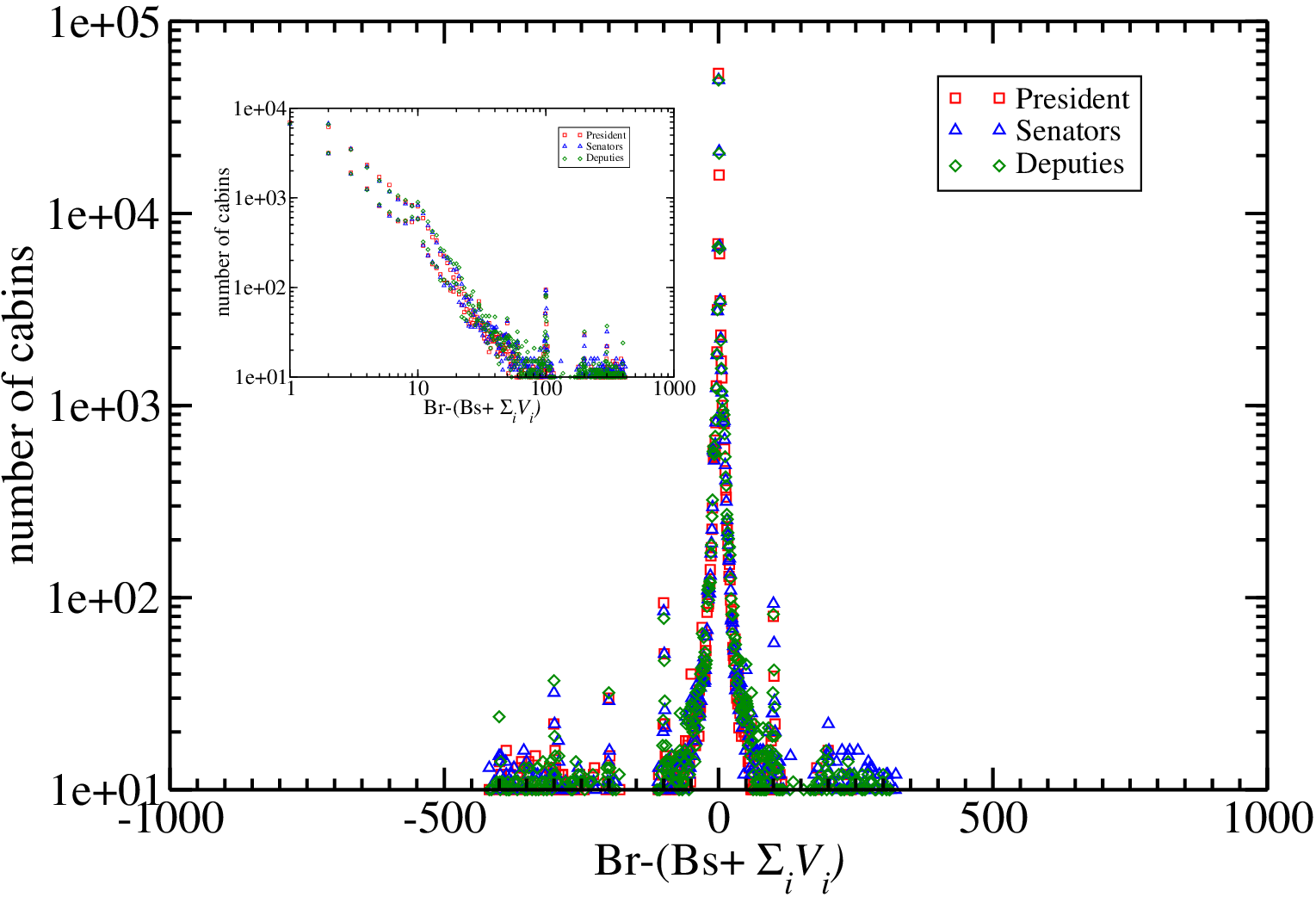}
\caption{(Color on-line) Error distributions, in logarithmic scale, for the difference between the total 
ballots received Br minus the sum of the ballots remaining Bs plus 
the sum of votes obtained by each political party, including null votes 
$\sum_i$ V$_i$, that is E$_3 = $ Br-(Bs+$\sum_i$ V$_i$). The meaning of the right and the left distribution 
branches is similar to that in figure \ref{Fig:LostBallots}.  
Notice that the  three distributions are similar. The inset shows both 
branches of the distributions in log-log scale.} 
\label{Fig:LostVotes}
\end{figure}

We start the analysis with E$_1$, that is, the total number of ballots, Br, against the sum
of the number of remaining ballots, Bs, and  the number of voters per cabin, V, i.e
., E$_1 =$ Br-(Bs+V).
As seen in 
Fig.~\ref{Fig:LostBallots}, the distribution of E$_1$ is 
peaked around zero but is neither the expected Dirac delta function $\delta(x)$, nor
a Gaussian nor a Lorentzian.
Data on the positive axis mean appearance of ballots, meanwhile data on the negative 
axis indicate lost of them. In plain English, in the latter case there are more 
registered people who voted than documents that certified their existence.  
Zooms in different parts of this distribution show the following 
facts: 
(1) The PREP preserve this number for only $\sim 45\%$ 
of the cabins. This result is unfortunate for the electoral authorities (IFE) since 
it says that PREP reliability is less than $50\%$. This is even worse since, as we shall
see later, the same occurs  for the rest of the error distributions considered and
in the distributions for both chambers.
(2) The distribution is not completely symmetric. In particular 
the peak around $-250$ is higher than the peak at $250$. 
(3) There are inconsistencies larger than $750$ votes for several 
cabins.  
(4) There are peaks at $\pm 10, \pm 20, \dots$ and also at 
$\pm 100, \pm 200, \dots$. These peaks, we assume, are related with 
capture typos, but it is not proved. 
(5) The peak at the left side of the distribution shows a 
different behavior for senators than the other cases. 
%This result cannot be statistically understood, since all 
%certificates (for president, senators and deputies) 
%are captured in the same way. 
%{\it A result like this means that the capture of the data was 
%different for three similar processes.} 
(6) The distribution between $1$ and $100$ decays as a power law as 
can be seen in the insert of figures~\ref{Fig:LostBallots} to \ref{Fig:LostVotes}.

In Figs.~\ref{Fig:LostDeposited} and~\ref{Fig:LostVotes} we show the distributions of  E$_2$ and E$_3$ 
conservation laws. 
%by plotting the distributions of differences between the total 
%ballots received Br and the sum of the ballots remaining Bs plus the 
%ballots deposited in urns Bd per cabin, i.e., Br-(Bs+Bd), for the first figure. In the second figure we plot 
%the difference between the total 
%ballots received Br and the sum of the ballots remaining Bs plus 
%the sum of votes obtained by each political party, including null votes 
%$\sum_{i}^{N}$ V$_i$, i.e., Br-(Bs+$\sum_i$ V$_i$).
Both distributions should give a delta function $\delta(x)$ 
in the ideal case. But, as seen in Figure~\ref{Fig:LostDeposited}, it is very asymmetric and present 
extreme values around $750$ as in Fig.~\ref{Fig:LostBallots}. In the insets we 
show that typos are also present and that power laws appear. The results for E$_4$ and E$_5$ are similar.

In the previous analysis all the cabins were considered, notwithstanding the fact that a selection 
is required, since many records present alterations as was documented by Crespo~\cite{Crespo}. 
However, statistical analysis, like the present one  are valid, 
since the universal source of error must be studied for future references.
Other errors, random or systematic,
require future analysis as well.

The result of {\it presidential} 2006 election  caused 
a debate about the existence or not of a Mega-fraud against the candidate of PRD and this 
section offers a new insight into the question. Since the error distributions of {\it all}
the cases have a similar behavior, we have only two options: i) There was no large fraud in presidential 
election, or ii) There was a large fraud in {\it all} the cases, including both chambers.

\begin{figure}[t!]
\includegraphics[width=1.0\columnwidth]{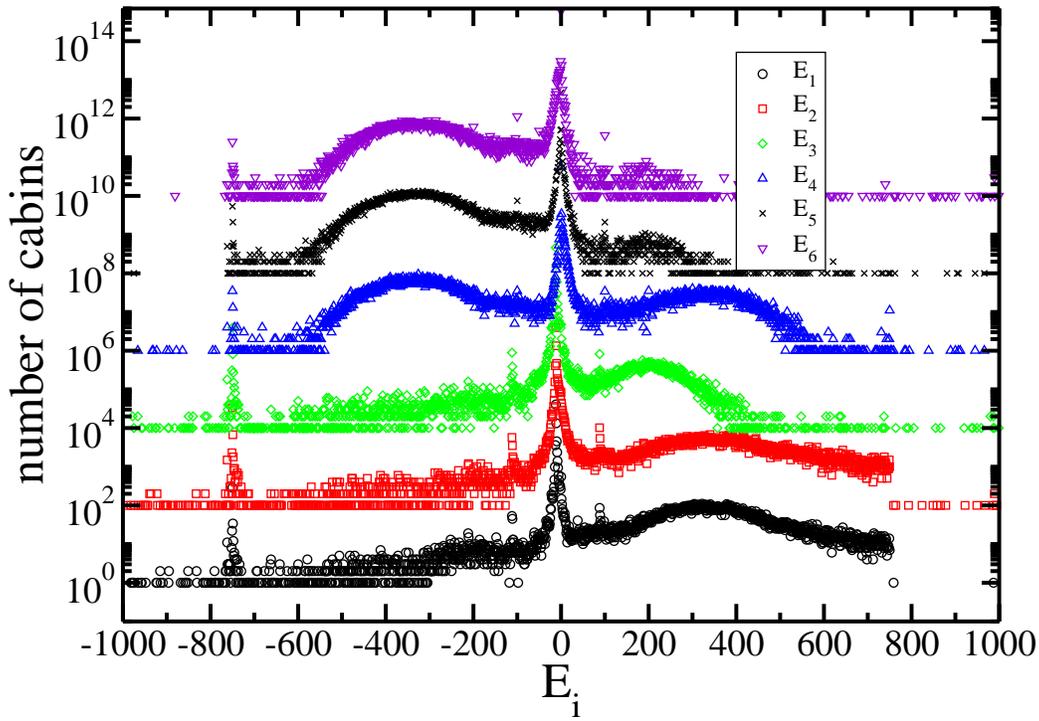}
\caption{(Color on-line) All error distributions for the 
presidential election in 2000. The meaning of E$_i$ are reported in 
Table~\ref{tabla2}. Date were taken from the corresponding
PREP file. In order to see all the cases in the same plot we 
multiply resulting distribution by a factor of one hundred each time.
As seen in the plot the global behavior is similar to those
appeared in the whole 2006 electoral process. The scale in $y$-axis 
is logarithmic.
}
\label{Fig:2000elec}
\end{figure}

In order to differentiate between the previous options we calculated the same 
error distributions for the PREP corresponding to the presidential election in 
2000. We summarize the results in Fig.~\ref{Fig:2000elec}. As can be seen, the main characteristics 
of whole election in 2006 are present in the 2000 election. Nobody mentioned 
a fraud or a mega-fraud then, but certainly the present
results tell something about the dynamics in Mexican elections. The large amount 
of errors and vicious practices certainly are present in both elections, 
and requires of a separate analysis. It is important to remark that it is not
only the amount of errors important, the {\it distributions} show the dynamics
and hopefully an analysis of them will provide us a better understanding of the
human processes that an election imply. Aparicio~\cite{Aparicio06a} reports that the percentage of errors
are similar in the presidential elections in 2000 and 2006, but here we offer
new tools to understand such a behavior. For instance, even when the amount of
received ballots at each cabin is around 750, the reported quantity present large
deviations, many of them, probably,  due to carelesnes of the citizen authorities record.
If this is of a malicious origin is unclear but opens the opportunity for new analysis. 
Another point is that performing sums under pressure could be a no trivial matter. 
Measuring the distribution of such behaviors is an open task to social scientists and 
suggest the performance of experiments as well.
allows to perform experiments as well.
For example, to design an experiment to obtain the distribution of typos under pressure in the 
capture line. To the knowledge of the authors this distribution has not been reported in the
literature. The meaning of the lobes with maximum around $\pm 250$ must be analyzed
as well as the power law decay in all the studied cases.   

As a final remark of this section, we note that with the large amount of errors
the result in the presidential election of 2006 could remain unclear and a 
technical tie.  However, analysis similar to the
presented here, including the chambers election and comparison with other 
electoral but equivalent processes, is useful in order to clarify the forensics of
fraud or not in future elections, in Mexico or in any democracy.

\section{Distribution of the votes}\label{Section:Distributions}

\begin{figure}[tb!]
\includegraphics[width=1.0\columnwidth]{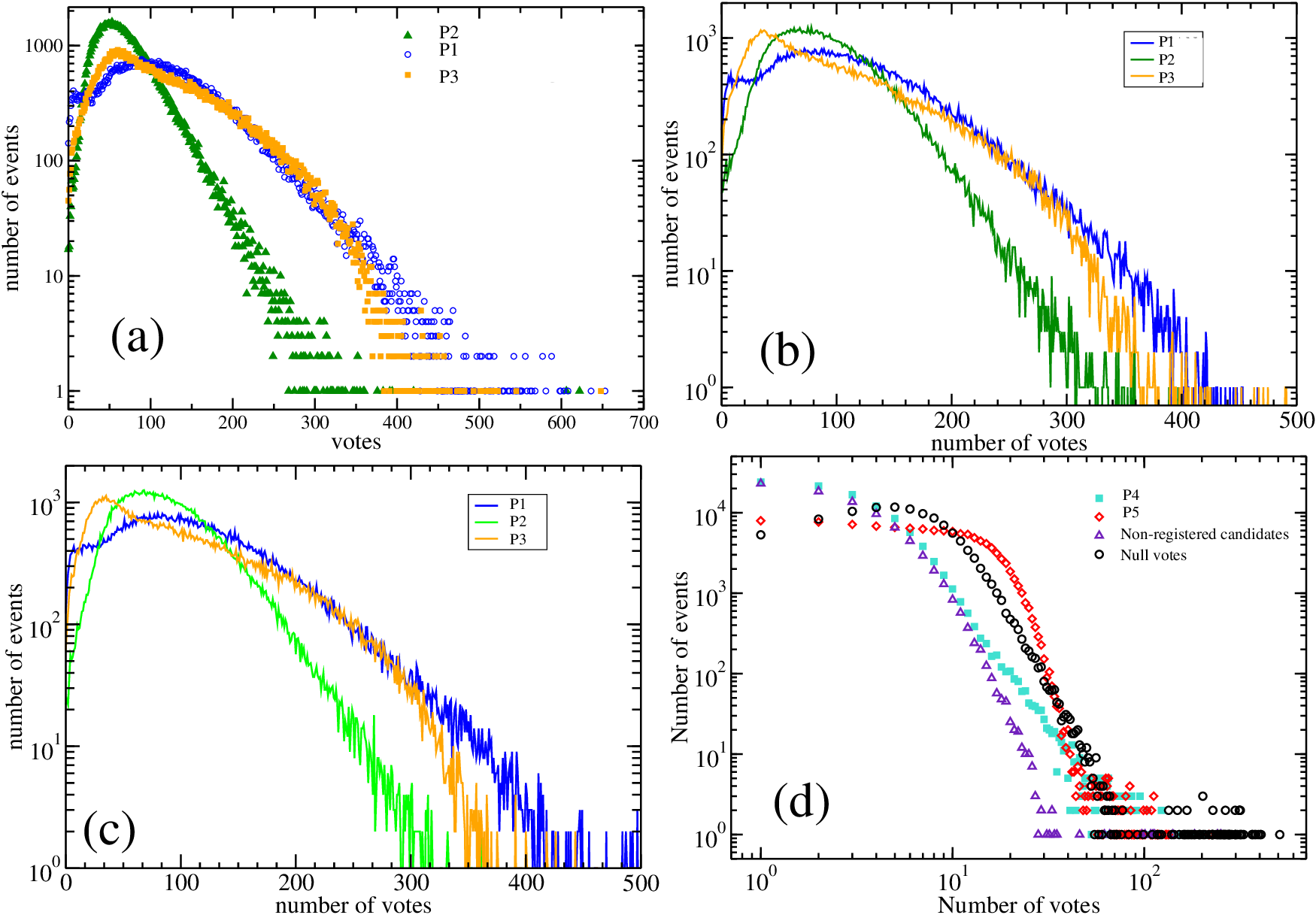}
\caption{Histograms showing the number of electoral cabins 
whith given number of votes for the three main parties, 
(a) president, (b) deputies,  (c) senators. In (d) the results votes 
for president by the small parties are given. Note that the P2 
distribution have an exponential tail while the distribution of the 
small parties show a shifted power-law.} 
\label{Fig:Histogram}
\end{figure}

As a final study we deal with the
histograms of the number of electoral cabins (polling stations) with a certain 
number of votes shown in Fig.~\ref{Fig:Histogram}. These histograms 
are the non normalized vote distribution for each party/candidate. The results 
for PAN,PRI and PRD presidential candidates  appear in Fig.~\ref{Fig:Histogram}(a) 
and the corresponding results for deputies and senators in panels (b) and (c).
Panel (d) of this figure corresponds to PANAL, Alternativa, non-registered candidates 
and null votes for the presidential election.

\begin{figure}[tb]
\begin{center}
\includegraphics[width=1.\columnwidth,angle=-90,scale=0.5]{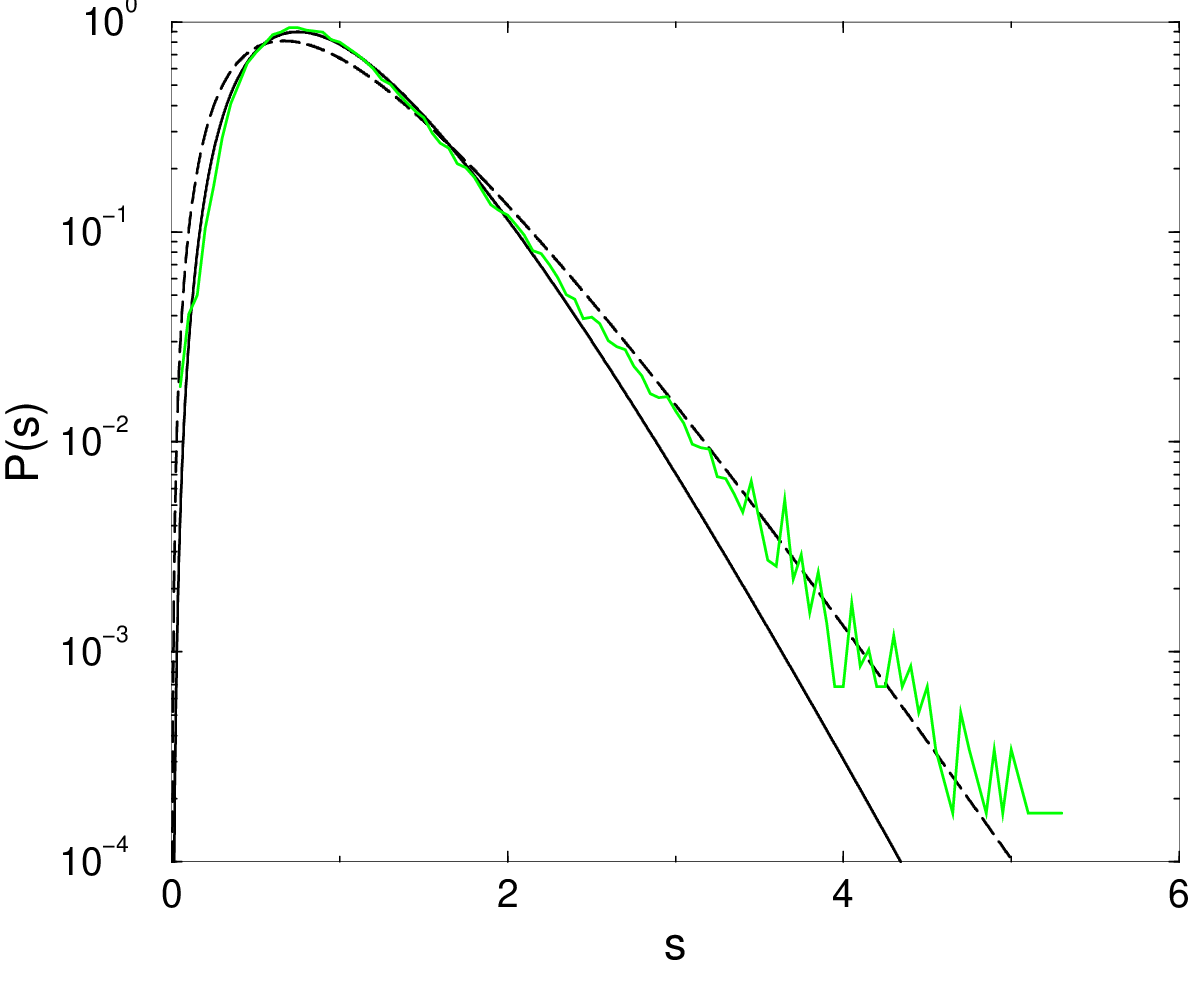}
\caption{Unfolded distribution for a randomly shuffled sequence of votes 
for the P2 presidential candidate (green line). A daisy model of 
rank $r=2$ and $3$ (see
text for details) are shown in dashed and continuous black lines.
} 
\label{Fig:PRIDistribution}
\end{center}
\end{figure}

Histograms  for the votes of PRI change very slowly in the three cases 
(Fig.~\ref{Fig:Histogram}(a) to (c)). 
The tail of these distributions looks exponential
(a fit is shown in Fig.~\ref{Fig:PRIDistribution}).  
This is not the case for the distributions of the two main presidential candidates 
(Fig.~\ref{Fig:Histogram}(a)).
The distribution of the votes for FCH/PAN shows a very different 
behavior for electoral cabins with less than $\sim 40$ votes, since it
starts flat. The distribution for AMLO/PRD is 
also irregular. It shows three different regimes. It appears like a 
distribution in which realizations between 60 and 300 votes are missing. 
This could be due to two reasons: the first, is that the data were manipulated; 
the second, is that the distribution of the votes for PRD is composed of two or more distributions 
corresponding to several voter dynamics~\cite{Merlin}. 
As an example, a weighted sum of two distributions
\begin{equation}
P = p P_{Daisy}(\langle v_1 \rangle,\sigma_1)+(1-p) P_{log-normal}(\langle v_2 \rangle,\sigma_2)
\end{equation}
with different centroids, $\langle v_1 \rangle$ and $\langle v_2 \rangle$, like a Daisy model (see below) and a 
log-normal distribution could give place to such deformed distributions. Both functions correspond to different groups
of voters, the first one to corporate voters
~\cite{HernandezSaldanaE1} 
and the other to 
certain proportional elections \cite{Castellano2007}. The distributions for the 
senators and deputies present similar behavior as seen in 
Fig.~\ref{Fig:Histogram} (b) and (c). In fact, the participation of a uniform 
group of voter in important elections is, a priori, unlikely; so, it is enough to 
have two kind of voters following distribution with different maxima to have such 
a behavior. This topic is matter of current research by one of us.

The histograms for the parties with a small number of votes are given in 
Fig.~\ref{Fig:Histogram}(d). As seen, all histograms have a similar behavior, except for a small numbers of votes. 
All are shifted power-laws, except for the PANAL presidential candidate which 
presents deviations. The results for deputies and senators for the same parties are similar (not
shown). These results
can be explained with several models of cluster growth
in complex networks~\cite{CostaFilho2006,Castellano2007}, for instance, 
and appear in other electoral processes~\cite{CostaFilho1999,Herrmann} for proportional votes. 
All these stidies are in search of a universal dynamics in voting processes.
A research along these lines for the Mexican elections is in progress. 
We found an inconsistency in the tail of the annulled votes.
This distribution shows several electoral cabins with more than 100 annulled 
votes. The probability of having such results is negligible, so that these 
results are not statistical, i.e., they are the result of negligence or of a 
malicious act. 

Finally, we return to the PRI case. The distribution of votes 
for this party  is  clearly smooth and a fit using the 
so called Daisy models~\cite{HernandezSaldana}
was performed previously on the 
final records (Distrital count) for Mexican elections in 2000, 2003 and 2006~\cite{HernandezSaldanaE1}.
The corresponding fit for the presidential PREP data corresponds to 
a rank $r =3$ model
\begin{equation}
P_3(s) = \frac{4^4}{3!} s^3 \exp(-4s),
\end{equation}
for the distribution's central part, and an $r=2$ model 

\begin{equation}
P_2(s) = \frac{3^3}{2!} s^2 \exp(-3s),
\end{equation}
for the tail. A plot on the normalized distribution of votes appears in 
Figure~\ref{Fig:PRIDistribution}. Here the ``unfolding'' of 
votes is performed using a $4$th degree polynomial on windows of $300$ and
$3000$ cabin certificates obtaining a reliable average density of events
\footnote{The ``unfolding'' procedure and their meaning is explained in 
\cite{HernandezSaldanaE2} but is of standard use in data analysis 
long ago.}. 
Notice that no fitting parameter was used, in contrast to 
a Brody or a Weibull  distributions. The reason of this behavior is unclear 
and requires a local analysis of the distribution in order to 
separate the states and municipalities  ruled traditionally by PRI.
A connection with a geographical approach could be of interest \cite{SuarezAlberro}

\section{Conclusions}

We studied some statistical properties of the federal
Mexican elections using the Program of Preliminary Electoral Results (PREP) database from 2000 and 2006.
Our main goal was to test reliability in electoral processes, mainly in a controversial
election where people suspect of a Mega-fraud. Since we do not have previous experience 
in electoral data and we do not know how specifically a Mega-fraud must look we
chose to test the hypothesis:

$H_1:$ {\sl Data from PREP for the Mexican elections of 2006, presidential and both chambers, 
present the same qualitative statistical behavior.} 

And the alternative one:

$H_0:$ {\sl Data from PREP for the Mexican elections of 2006, presidential and both chambers,
present a qualitative different statistical behavior.}

We did tried our best effort to test these hypothesis. 

The first suspicion of cheating behavior was the non crossing between the vote
percentage of AMLO and FCH in the real time data. We demonstrate that the 
percentage of votes are correlated while the number of votes are not. 
The change in slope is strongly altered by the PRI increase of the number of votes around 
the $70\%$ of the processed vote certificates. This behaviour is reproduced in both chambers 
election of the same year and in the presidential election  in 2000.
This contradicts hypothesis H$_0$ for the time behavior.

A second test concerns the self consistency of the data-set that appears in the PREP file. For an ideal case the 
sum of votes for all party/candidates and the remaining  ballots must be equal to the number of ballots received 
at the poll station in the election day, for instance. From the set of data in the file six different and 
independent test of self consistency can be constructed (see \ref{tabla2} ). All of them can be subject 
of error, from a honest human mistake to a cheating alteration in many of the known ways.
Hence we built up the error distribution, which consist of the histogram of the selected error 
for all the cabins records. We found a similar global behavior in all the cases, for the 
presidential election in 2006, which is under suspicion of fraud, for both chambers election 
in the same year and, in the presidential election in 2000. The later was assumed as a clean or at least not disputed
election for the participants. In all the cases, the error distributions have a central peak with 
a power law decay in, approximately, the interval from $-50$ to $50$. All of them have a 
lobe around $\pm300$ or $\pm400$, but generally asymmetric. Some revivals or peaks
in $\pm10$, $\pm 20$,$\dots$, appear in the 2006 cases and they are less prominent in 
the presidential election of 2000. We suspect of typos for this kind of errors, but, 
as all the characteristics presented in the current paper, these must be tested with experiments 
and with new evidence. For instance, at our best knowledge, there is no  distribution 
of typos under pressure conditions. Hence for the self consistency test no large 
deviations were observed and, H$_1$ is confirmed.

%We have shown that the appearance of the data in real  time are not 
%distributed in a random way, and that this depends on the way the sample is taken. 
%Evidence of correlations between parties with large numbers of
%votes and small numbers of votes is evident. There is also evidence of correlation
%with the annulled votes. Quantities that should be conserved, for self-consistency of the 
%records, were studied. The distribution of such quantities  have two main behaviors, 
%with a power law in the central part instead of the expected Gaussian or Lorentzian distribution.
%Unexplained peaks are present in all the distributions (see figures 
%\ref{Fig:LostBallots}-\ref{Fig:LostVotes}.)
%The number of records with inconsistencies is around $50\%$ in all cases, for
%the president election and for the upper and lower chambers as well.
%In reference to the existence of a major fraud against P3 candidate we offer a new 
%insight. Since the error distributions of {\it all}
%the cases have a similar behavior, we have only two options:
Even with the evidence provided here and in the peer reviwed literature 
it is possible the existence  of a large  fraud, but the probability of 
such a case is small. Clarification of July 2006 presidential election in Mexico is
of a historical importance and hence requiere of more analysis. 
What we consider much more interesting at this is to 
insist that electoral processes are subject to a lot of 
errors and comparisons with theoretical models and predictions
must take this fact in to account. We offered, that is our hope,  new ways 
to tipify errors. It is not enough to talk about the 
magnitude of the errors, since a dynamics inside exists. We leave to an ulterior
work comments on the way the cheating could appear in the error distributions. 
For instance, another explanation for the peaks in such a distributions
is that the party representative altered the vote record by adding a zero to the 
total votes obtained by their own party, transforming  a $7$ into $70$, for instance.
Or how the errors are distributed along the maginalization index?. Are they
distributed in the same way or are different?. How to the different and well 
know malicious behaviour apperars in these distributions?

% i) There was no large fraud in presidential
%election, or ii) There was a large fraud in {\it all} the cases, including both chambers.

As a final remark, we also have obtained the distributions of votes for the different parties. 
In particular, the distribution of the party that was in power in Mexico 
during more than $70$ years behaves smoothly. Daisy models of 2nd and 3rd rank seem 
to fit different parts of the measured distribution. In contrast, 
the distributions of the parties with more votes are more complex and, probably,
composed of different voters dynamics.
Distributions of small parties follow power laws. This behavior  is consistent 
with several theoretical models.  
%The difference 
%between the first and second place should be larger than the error 
%associated to the measurement, in this case, the electoral processes.

\section{Acknowledgments}
This work was supported by DGAPA-UNAM, PROMEP/SEP and 
projects UAM-A CBI. We  
thank C. Badillo for useful comments. We wish also thank  E. Morf\'{\i}n for 
his invaluable help with the database of the PREP, and  A. Baqueiro 
for allowing us the use of real-time captured data of the PREP for Fig.~1.

\section{Appendix: Additional data}

In this appendix we present additional data, required for comparison. In tables
\ref{tabla3} and \ref{tabla4} appear the correlations for percentage of vote 
(lower matrix) and for number of votes (upper matrix). The correlations were 
calculated using the standard method
\begin{equation}
c(x,y) =\frac{1}{N}\sum^N_i \hat{x}(i) \hat{y}(i),
\end{equation}
where 
\begin{equation}
\hat{x}(i) = (x(i) - \bar{x})/\sqrt{Var(x)},
\end{equation}
for the corresponding variable, the number of votes or the percentage of them, for all
the combinations of political parties.

\begin{figure}[h]
\begin{center}
\includegraphics[width=1.\columnwidth,angle=0,scale=0.9]{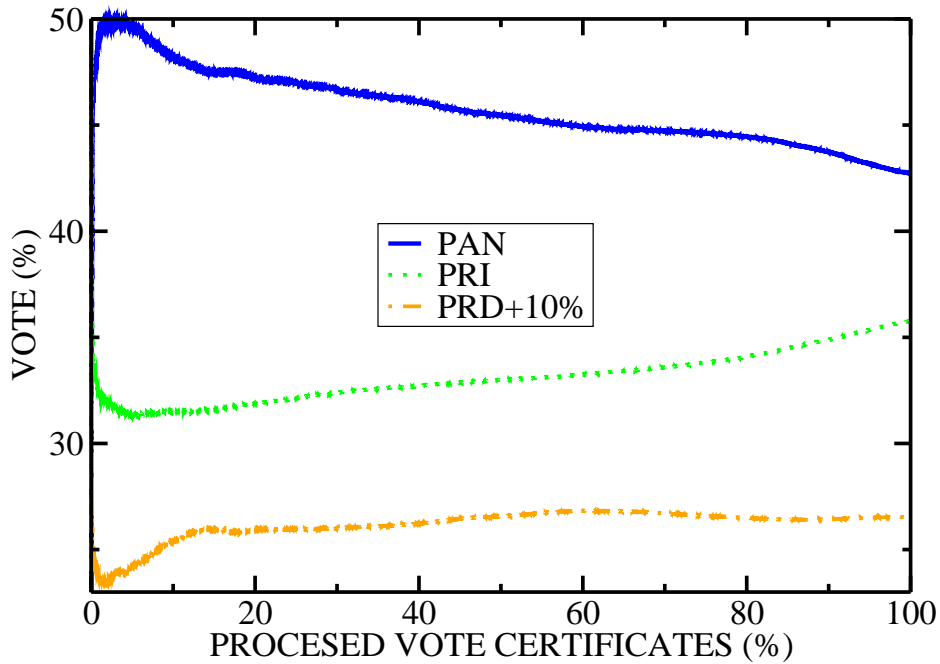}
\caption{(Color on-line) Percentage of votes as a function of processed 
vote certificates for the July 2000 presidential election. We use 
the PREP data and only the three main candidates are shown. Notice the
slope change for PRI candidate around $70\%$ of processed vote certificates.
} 
\label{Fig:PREP2000}
\end{center}
\end{figure}

In Figure \ref{Fig:PREP2000} the time behaviour of PREP file during July 2000 is shown.In this case, the PRI appears in second place, but with a change in its slope around 
$70\%$ of polling stations processed. Shifting these percentages in order to show them 
as in july 2000 presidential election resembles the same bahavior. The shifting is not shown. Notice that in this figure as well as in all in the paper consider all the data
available.

\begin{table}[ht]

\begin{center}
%\label{tabla3}
\begin{tabular}{|l||r|r|r|r|r|r|r|}
\hline
      &   PAN  & PRI  & PRD     &  PANAL & Alter & N.R.  & N.V.  \\ \hline \hline
PAN   &  1.00 &  0.06 &  -0.3   &  0.22  & 0.22  & 0.00  & -0.04      \\ %\hline
PRI   & -0.45 &  1.02 &  -0.19  &  0.02  & -0.18 & 0.00  &  0.13     \\ %\hline
PRD   &-0.82  & -0.16 & 1.02    &  0.18  &  0.41 & 0.05  &  0.03     \\ %\hline 
PANAL & 0.42  & -0.76 & 0.01     &  0.71  &  0.36 & 0.05  &  0.02     \\ %\hline
ALTER & 0.48  & -0.97 & 0.08    &  0.79   & 1.00 & 0.06  & -0.02      \\ %\hline
N.R.  & -0.84 &  0.22 & 0.78    & -0.26   & -0.27 & 0.90  & 0.02      \\ %\hline 
N.V.  &  0.91 &  0.42 & 0.72    & -0.52   & -0.51 & 0.81  & 1.00  \\ \hline

\end{tabular}
\protect\caption{\label{tabla3}Correlation matrix of {\it vote percentage}, lower triangle, and {\it number of votes}, upper triangle, for the deputies election in 2006. Diagonal corresponds to auto-correlation of vote percentage. N.R. stands for non registered candidates and N.V. for
annulled votes.}
\end{center}
\end{table}

\begin{table}[h] %\label{tabla4}

\begin{center}
\begin{tabular}{|l||r|r|r|r|r|r|r|}
\hline
      &   PAN     & PRI       & PRD       &  PANAL    & Alter      & I.C.       & N.V.   \\ \hline \hline
PAN   & 1.00 & 0.08      & -0.31     &  0.20     & 0.21       & 0.01       & -0.05      \\ % \hline 
PRI   & -0.62     & 1.02 & -0.22     &  -0.03    & -0.19      & -0.00      & 0.12     \\ %\hline
PRD   & -0.83     & 0.06      & 1.02 &  0.28     & 0.40       & 0.02       & 0.03      \\ %\hline
PANAL & 0.28      & -0.71     & 0.14      & 0.62 & 0.45       & 0.06       & 0.00      \\ %\hline
ALTER & 0.22      & -0.86     & 0.34      & 0.73      & 1.00  & 0.06      & -0.02      \\ %\hline
I.C.  & -0.50     & 0.49      & 0.27      & -0.32     & -0.38      & 0.76  & 0.02        \\ %\hline 
N.V.  & -0.93     & 0.62      & 0.73      & -0.36     & -0.32      & 0.53      & 1.00  \\ \hline
\end{tabular}
\protect\caption{\label{tabla4}Correlation matrix of {\it vote percentage}, 
lower triangle, and
 number of votes, upper triangle, for senatorial election in 2006. Diagonal corresponds to auto-correlation of vote percentage. N.R. stands for non registered candidate
s and N.V. for annulled votes.}

\end{center}
\end{table}

\end{document}